\documentclass[12pt,twoside,a4paper]{article}
\pdfoutput=1

\usepackage{amsmath, amsthm, amsfonts, amssymb}
\usepackage{graphicx}
\usepackage{multirow}
\usepackage{float}
\usepackage{color}
\usepackage{cite}
\usepackage[width=\textwidth,font=small,labelfont=bf,
labelsep=endash]{caption}
\usepackage{titlesec}

\titleformat{\section}
  {\normalfont\fontsize{13}{16}\bfseries}{\thesection}{1em}{}
\titleformat{\subsection}
  {\normalfont\fontsize{11}{14}\bfseries}{\thesubsection}{1em}{}

\normalsize
\newcommand\tr            {\mathrm{Tr}}
\newcommand\re            {\mathrm{Re}}
\newcommand\im            {\mathrm{Im}}

\numberwithin{equation}{section}
\makeatletter
\newcommand{\vast}{\bBigg@{4}}
\newcommand{\Vast}{\bBigg@{5}}
\makeatother

\begin{document}

\title{\textbf{Entanglement in harmonic systems at coherent states}}
\author{Dimitrios Katsinis$^{1,2,3}$ and Georgios Pastras$^{2,4}$}
\date{$^1$Instituto de F\'isica, Universidade de S\~ao Paulo, Rua do Mat\~ao Travessa 1371, 05508-090 S\~ao Paulo, SP, Brazil\\
$^2$NCSR ``Demokritos'', Institute of Nuclear and Particle Physics,\\15310 Aghia Paraskevi, Attiki, Greece\\
$^3$National and Kapodistrian University of Athens,\\ Department of Physics, 15784 Zografou, Attiki, Greece\\
$^4$Laboratory for Manufacturing Systems and Automation, Department of Mechanical Engineering and Aeronautics,\\ University of Patras, 26110 Patra, Greece
\linebreak\\
\texttt{pastras@lms.mech.upatras.gr dkatsinis@phys.uoa.gr}}

\vskip .5cm

\maketitle

\begin{abstract}
It is well-known that entanglement entropy in field theory at its ground state is dominated by an area law term, presenting a similarity to the entropy of black holes. It is interesting to investigate whether this similarity can be extended by showing that gravitational dynamics emerges from the first law of entanglement thermodynamics. Answering this question requires the specification of the modular Hamiltonian. Motivated by the above, we study entanglement in the toy model of harmonic systems lying at any classicalmost state, i.e. any coherent state. We specify explicitly the reduced density matrix and its time-evolution, as well as the modular Hamiltonian. Interestingly, the time evolution is unitary and we specify the effective Hamiltonian which generates it. Our results provide the tools to investigate the similarity between gravity and entanglement in discretized free scalar field theory in the framework of \cite{srednicki}.
\end{abstract}

\newpage

\tableofcontents

\newpage
%\DeclareMathOperator{\arccosh}{arccosh}
%\DeclareMathOperator{\arccot}{arccot}
%\DeclareMathOperator{\arctanh}{arctanh}
%\DeclareMathOperator{\arccoth}{arccoth}
%\DeclareMathOperator{\csch}{csch}
%\DeclareMathOperator{\tr}{Tr}
%\DeclareMathOperator{\re}{Re}
%\DeclareMathOperator{\im}{Im}
%%\usepackage{bm}
%\title{}
%\author{Dimitris Katsinis, Georgios Pastras}
%\begin{document}
%\maketitle 
%\newpage
%\tableofcontents
%\newpage
\section{Introduction}

In 1993, in a seminal paper, Srednicki \cite{srednicki} showed that the entanglement entropy in free massless scalar field theory at its ground state is proportional to the area of the entangling surface. This apparent similarity of entanglement entropy and black hole entropy gave rise to the question about whether black hole entropy can be attributed to quantum entanglement between the interior and the exterior of the black hole. Such a scenario would favour a description of gravity as a quantum entropic force \cite{VanRaamsdonk:comments,VanRaamsdonk:building}.

The latter hypothesis is greatly supported by the holographic duality. In any composite quantum system, the first law of entanglement thermodynamics trivially holds. This relates the variation of entanglement entropy and the mean value of the modular Hamiltonian\footnote{The modular Hamiltonian is defined as the logarithm of the density matrix $H_{\mathrm{mod}} = - \ln \rho$} under a variation of the state of the composite system
\begin{equation}
\delta S_{\mathrm{EE}} = \delta \left< H_{\mathrm{mod}} \right> .
\end{equation}
This relation is true for the holographic realizations of the entanglement entropy and the expected value of the modular Hamiltonian for spherical entangling surfaces in the bulk, if and only if the small variation of the state in the boundary corresponds to a small variation of the geometry in the bulk, which obeys the linearized Einstein equation \cite{Lashkari:2013koa}. Therefore, the holographically emergent gravity is at least not contradictory to a quantum entropic gravitational force scenario.

Further investigation of the relation between gravity and quantum statistics attributed to quantum entanglement requires the specification of the modular Hamiltonian for non-trivial states of the underlying theory. This is an extremely difficult task, since it requires the explicit calculation of the reduced density matrix and not only its spectrum. In this manner, the Replica trick and holographic calculations of relevant quantities, such as the entanglement entropy, are disadvantageous, since they naturally lead to the calculation of traces of integer powers of the reduced density matrix. This data are adequate to reproduce in principle the spectrum of the latter but not its eigenstates. On the other hand, the simplistic, brute force approach of \cite{srednicki}, although it applies to the non-realistic, non-interacting integrable toy model of free scalar field theory, it has the advantage that it directly calculates the reduced density matrix, and, thus, it can lead to the explicit calculation of the modular Hamiltonian operator.

In this work, we provide the tools to extend the study of the similarity between gravity and entanglement in the framework of harmonic systems, as initiated in \cite{srednicki}, calculating the modular Hamiltonian, as an operator, which is a function of the positions and momenta of the local degrees of freedom, and specifying the time evolution of the reduced system. This task is performed for a general classicalmost state of the composite system, i.e. a coherent state. In section \ref{sec:Osc}, we develop the algebraic solution to a generalized oscillator. This system is described by a Hamiltonian, which is a quadratic function of both position and momentum, but with explicit time dependence. This generalized oscillator is going to be the basic building block of the modular Hamiltonian. In section \ref{sec:two} we find the modular Hamiltonian and specify the time evolution of the reduced system in the simplest problem of two coupled oscillators. In section \ref{sec:many}, we extend the results of section \ref{sec:two} to an arbitrary harmonic system. Finally, in section \ref{sec:discussion} we discuss our results. There is also an appendix, containing an application of the generalized harmonic oscillator that we developed in section \ref{sec:Osc}, in the specification of the most general Gaussian states of the simple harmonic oscillator.

\section{Generalized Harmonic Oscillator}
\label{sec:Osc}
In this section we generalize the usual algebraic solution of the simple harmonic oscillator to construct a more general quadratic Hamiltonian, which has explicit time-dependence. This problem has been considered a long time ago \cite{Husimi_osc,closed_form_osc} from a completely different perspective. The time-dependent Schrödinger equation has a tower of solutions, which resembles the Fock space of the simple harmonic oscillator. More specifically, the lowest state of this tower is a time-dependent Gaussian state. The entanglement in harmonic systems at coherent states, which are in fact time-dependent Gaussian states, is naturally described with the use of this tool.

The problem can be treated algebraically in a similar manner to the simple harmonic oscillator \cite{Meyer_osc,Ballhausen,moller_jorgensen}. Let us recall the main idea behind the algebraic solution of the simple harmonic oscillator. The ground state is a Gaussian state, with equal uncertainties in position of momentum. It is a trivial fact that any Gaussian state is annihilated by an appropriate linear combination of the position and momentum operators. This linear combination is what we call the annihilation operator $a$; its Hermitian conjugate $a^\dagger$ is the so-called creation operator. It turns out that the commutator of these two operators is equal to the identity operator and furthermore the Hamiltonian $H$ of the harmonic oscillator assumes the form $\hbar \omega \left( a^\dagger a + 1 / 2 \right)$. This implies the well-known commutation relations $\left[ H , a \right] = \hbar \omega a , \left[ H , a^\dagger \right] = \hbar \omega a^\dagger$, which provide the spectrum and eigenstates of the harmonic oscillator.

We want to generalize this procedure, but instead of the ground state of the harmonic oscillator, we will consider the most general Gaussian state with time-dependent parameters. For this reason, in our construction, we consider as annihilation and creation operators $A$ and $A^\dagger$, the most general linear combinations of the position and momentum operators, which satisfy the commutation relation
\begin{equation}
\left[ \hat{A} , \hat{A}^\dagger \right] = 1 .
\end{equation}
It is trivial to show that these operators read
\begin{align}
\hat{A} &:= \frac{1}{\sqrt{2 m \hbar \omega^R}} \left[i \left( \hat{p} - p_0 \right) + m\left( \omega^R + i \omega^I \right) \left(\hat{x} - x_0 \right) \right] , \label{eq:annihilation}\\
\hat{A}^\dagger &:= \frac{1}{\sqrt{2 m \hbar \omega^R}} \left[- i \left( \hat{p} - p_0 \right) + m \left( \omega^R - i \omega^I \right) \left( \hat{x} - x_0 \right) \right] , \label{eq:creation}
\end{align}
where $x_0$, $p_0$, $\omega^R$ and $\omega^I$ are arbitrary functions of time. For the sake of simplicity, we assume that the parameter $m$ is constant; notice that this parameter appears always in the form of the product $m \omega^R$ or $m \omega^I$, and, thus its time dependence can be absorbed in the time dependence of the parameters $\omega^R$ and $\omega^I$. One could also add a global time-dependent phase to these operators, but this is physically indifferent.

It is straightforward to show that the normalized state, which is annihilated by $\hat{A}$, reads
\begin{equation}
\left\langle x \middle| \Phi_0 \right\rangle = \left( \frac{m \omega^R}{\pi \hbar} \right)^{1/4} \exp \left[ - \frac{m \left( \omega^R + i \omega^I \right)}{2 \hbar} \left( x - x_0 \right)^2 + i \frac{p_0}{\hbar}\left( x - x_0 \right) - \frac{i}{2} \varphi \right] ,
\label{eq:Phi_0}
\end{equation}
where $\varphi$ is in general an arbitrary function of time.

In a trivial manner, the states
\begin{equation}
\left| \Phi_n \right\rangle := \left( \hat{A}^\dagger \right)^n \left| \Phi_0 \right\rangle
\end{equation}
are eigenfunctions of the operator
\begin{equation}
\hat{\tilde{H}} := \hbar \omega^R \hat{A}^\dagger \hat{A} ,
\end{equation}
with eigenvalues equal to
\begin{equation}
\hat{\tilde{H}} \left| \Phi_n \right\rangle = n \hbar \omega^R \left| \Phi_n \right\rangle .
\end{equation}
When expressed in terms of the position and momentum operators, the operator $\hat{\tilde{H}}$ assumes the form
\begin{multline}
\hat{\tilde{H}} = \frac{1}{2m} \left( \hat{p} - p_0 \right)^2 + \frac{1}{2} \omega^I\left[ \left( \hat{x} - x_0 \right) \left( \hat{p} - p_0 \right) + \left( \hat{p} - p_0 \right) \left( \hat{x} - x_0 \right) \right] \\
+ \frac{1}{2} m \left[ \left( \omega^R \right)^2 + \left( \omega^I \right)^2 \right] \left( \hat{x} - x_0 \right)^2 + \frac{\omega^R}{2} .
\label{eq:AdagA_x_p}
\end{multline}

Since the operators $A$ and $A^\dagger$ contain an explicit time dependence, the operator $\hat{\tilde{H}}$ cannot emerge as the Hamiltonian in a time-independent Schrödinger equation; The states $\Psi_n$ are not solutions of the time-dependent Schrödinger equation $i \hbar \partial_t \Psi = \hat{\tilde{H}} \Psi$.

In order to upgrade the states $\Phi_n$ to solutions of a time-dependent Schrödinger equation, we need to find a time-dependent Hamiltonian $\hat{H}$, which obeys specific commutation relations with the creation and annihilation operators \eqref{eq:annihilation} and \eqref{eq:creation}, namely,
\begin{align}
i \hbar \dot{\hat{A}} = \left[ \hat{H} , \hat{A} \right] + \hbar \omega^R \hat{A} , \qquad i \hbar \dot{\hat{A}}^\dagger = \left[ \hat{H} , \hat{A}^\dagger \right] - \hbar \omega^R \hat{A}^\dagger .
\label{eq:A_evolution}
\end{align}
It is a matter of algebra to show that this Hamiltonian reads\footnote{Notice that had we considered a time-dependent mass $m$, this Hamiltonian would read
\begin{multline}
\hat{H} = \hbar \Bigg[ \left( \omega^R + \frac{\partial_t{\left(m\omega^I\right)}}{2m \omega^R} \right) \left( \hat{A}^\dagger \hat{A} + \frac{1}{2} \right) + \frac{\partial_t{\left(m\left(\omega^I + i\omega^R\right)\right)}}{4m\omega^R} \hat{A}^{2} + \frac{\partial_t{\left(m\left(\omega^I - i\omega^R\right)\right)}}{4m\omega^R} \hat{A}^{\dagger 2} \\
- \frac{\dot{p}_0 + m \left( \omega^I - i \omega^R \right) \dot{x}_0}{\sqrt{2m \hbar \omega^R}} \hat{A}^\dagger - \frac{\dot{p}_0 + m \left(\omega^I + i \omega^R \right) \dot{x}_0}{\sqrt{2 m \hbar \omega^R}} \hat{A} + \frac{p_0 \dot{x}_0}{\hbar} \Bigg].
\label{eq:H_A_Adagger_m_time}
\end{multline}}
\begin{multline}
\hat{H} = \hbar \Bigg[ \left( \omega^R + \frac{\dot{\omega}^I}{2 \omega^R} \right) \left( \hat{A}^\dagger \hat{A} + \frac{1}{2} \right) + \frac{ \dot{\omega}^I + i \dot{\omega}^R}{4 \omega^R} \hat{A}^{2} + \frac{\dot{\omega}^I - i \dot{\omega}^R}{4 \omega^R} \hat{A}^{\dagger 2} \\
- \frac{\dot{p}_0 + m \left( \omega^I - i \omega^R \right) \dot{x}_0}{\sqrt{2m \hbar \omega^R}} \hat{A}^\dagger - \frac{\dot{p}_0 + m \left(\omega^I + i \omega^R \right) \dot{x}_0}{\sqrt{2 m \hbar \omega^R}} \hat{A} + \frac{p_0 \dot{x}_0}{\hbar} \Bigg] .
\label{eq:H_A_Adagger}
\end{multline}
Expressing this Hamiltonian in terms of the position and momentum operators yields
\begin{multline}
\hat{H} = \frac{1}{2m} \left( \hat{p} - p_0 + m \dot{x}_0 \right)^2 + \frac{1}{2} \mathcal{A} \left[ \left( \hat{x} - x_0 \right) \hat{p} + \hat{p} \left( \hat{x} - x_0 \right) \right] + \frac{1}{2} m \tilde{\Omega}^2 \left( \hat{x} - x_0 \right)^2 \\
- \left( p_0 \mathcal{A} + \dot{p}_0 \right) \left( \hat{x} - x_0 \right) + \dot{x}_0 \left( p_0 - \frac{1}{2} m \dot{x}_0 \right) ,
\label{eq:H_x_p}
\end{multline}
where $\tilde{\Omega}^2$ and $\mathcal{A}$ are given by
\begin{equation}
\tilde{\Omega}^2 = \left( \omega^R \right)^2 + \dot{\omega}^I - \left( \omega^I \right)^2 + 2 \omega^I \mathcal{A} , \qquad \mathcal{A} = \omega^I - \frac{\dot{\omega}^R}{2 \omega^R} .
\end{equation}

The Gaussian state \eqref{eq:Phi_0} does not solve the Schrödinger equation
\begin{equation}
i \hbar \partial_t \left| \Psi \right\rangle = \hat{H} \left| \Psi \right\rangle
\label{eq:Schrodinger_time_dependent}
\end{equation}
for any function $\varphi \left( t \right)$. This function is determined by the Schrödinger equation. Plugging \eqref{eq:Psi_0} into \eqref{eq:Schrodinger_time_dependent} yields
\begin{equation}
\dot{\varphi} = \omega^R .
\end{equation}
The simplicity of this result is due to the specific choice of the $c$-number terms that we made in the definition of the Hamiltonian \eqref{eq:H_A_Adagger}. We define this solution of Schrödinger equation as the state $\left| \Psi_0 \right\rangle$,
\begin{equation}
\left\langle x \middle| \Psi_0 \right\rangle = \left( \frac{m \omega^R}{\pi \hbar} \right)^{1/4} \exp\left[ - \frac{m \left( \omega^R + i \omega^I \right)}{2 \hbar} \left( x - x_0 \right)^2 + i \frac{p_0}{\hbar}\left( x - x_0 \right) - \frac{i}{2} \phi \right] ,
\label{eq:Psi_0}
\end{equation}
where
\begin{equation}
\phi := \int dt \omega^R .
\label{eq:time_phase}
\end{equation}

%It is a matter of algebra to show that
%
%We define a state $\vert0\rangle$ by
%\begin{equation}
%\hat{A}\vert 0\rangle=0,\qquad \langle 0 \vert0\rangle=1.
%\end{equation}
%Now let us consider the state
%\begin{equation}
%\vert\Phi\rangle=\left(i\hbar\partial_t-\hat{H}\right)\vert 0\rangle.
%\end{equation}
%It follows that
%\begin{equation}
%\hat{A}\vert\Phi\rangle=-\left(i\hbar\dot{\hat{A}}-\left[\hat{H},\hat{A}\right]\right)\vert 0\rangle\propto \hat{A}\vert 0\rangle=0,
%\end{equation}
%where we used equation \eqref{eq:A_evolution}. As a result, $\vert\Phi\rangle=c(t)\vert 0\rangle$, implying that the state $\vert 0\rangle$ satisfies the equation
%\begin{equation}
%i\hbar\partial_t\vert 0\rangle=\left(\hat{H}+c(t)\right)\vert 0\rangle.
%\end{equation}
%As the state $\vert0\rangle$ is normalized, $c(t)$ has to be real. Thus, $c(t)$ can be absorbed by introducing an appropriate phase to the wave-function, so that $\vert0\rangle$ is a solution of the Schrödinger equation. Let us be more specific. The form of $\hat{A}$ suggest that the wave-function $\Psi_0$ is Gaussian. 
%Plugging this wave-function in the time-dependent Schrödinger equation results in

Since we have a solution of the Schrödinger equation \eqref{eq:Schrodinger_time_dependent}, we can construct an infinite tower of solutions algebraically, in a similar manner to the simple harmonic oscillator. Assume that $\left| \Psi \right\rangle$ is a solution of the Schrödinger equation \eqref{eq:Schrodinger_time_dependent}.
%\begin{equation}
%i\hbar\partial_t\vert \Psi\rangle=\hat{H}\vert \Psi\rangle,
%\end{equation}
Then, the state
\begin{equation}
\left| \Psi^\prime \right\rangle = e^{- i \phi} \hat{A}^\dagger \left| \Psi \right\rangle ,
\end{equation}
where $\phi$ is given by \eqref{eq:time_phase}, is another solution of the Schrödinger equation. In particular, in view of equation \eqref{eq:A_evolution} it follows that
\begin{equation}
i \hbar \partial_t \left| \Psi^\prime \right\rangle = \hbar \omega^R e^{- i \phi}\hat{A}^\dagger \left| \Psi^\prime \right\rangle + i \hbar e^{- i \phi} \dot{\hat{A}}^\dagger \left| \Psi^\prime \right\rangle + e^{- i \phi} \hat{A}^\dagger \hat{H} \left| \Psi^\prime \right\rangle = \hat{H} \left| \Psi^\prime \right\rangle .
\end{equation}

Similarly to the case of the simple harmonic oscillator, we define the states $\Psi_n$ as,
\begin{equation}
\left| \Psi_n \right\rangle = \frac{e^{- i n \phi}}{\sqrt{n!}} \left( \hat{A}^\dagger \right)^n \left| \Psi_0 \right\rangle .
\end{equation}
These states obey
\begin{equation}
\hat{A} \left| \Psi_n \right\rangle = e^{- i \phi} \sqrt{n} \left| \Psi_{n - 1} \right\rangle , \qquad \hat{A}^\dagger \left| \Psi_n \right\rangle = e^{i \phi} \sqrt{n + 1} \left| \Psi_{n + 1} \right\rangle
\end{equation}
and
\begin{equation}
\hat{A}^\dagger \hat{A} \left| \Psi_n \right\rangle = n \left| \Psi_n \right\rangle .
\end{equation}
Recalling the form of the position representation of the wavefunctions of the simple harmonic oscillator, it is not difficult to show that, the position representation of these solutions reads
\begin{multline}
\left\langle x \middle| \Psi_n \right\rangle = \frac{1}{\sqrt{2^n n!}} \left( \frac{m \omega^R}{\pi \hbar} \right)^{1/4} H_n \left( \sqrt{\frac{m \omega^R}{\hbar}} \left( x - x_0 \right) \right) \exp \left[ - \frac{m \omega^R}{2 \hbar} \left( x - x_0 \right)^2 \right] \\
\times \exp \left[ - i \frac{m \omega^I}{2 \hbar} \left( x - x_0 \right)^2 + i \frac{p_0}{\hbar} \left( x - x_0 \right) - i \left( n + \frac{1}{2} \right) \phi \right] ,
\label{eq:wavefunctions_general}
\end{multline}
where $H_n$ is the Hermite polynomial of order $n$. This justifies the choice of $c$-number terms of the Hamiltonian \eqref{eq:H_A_Adagger}. They have been appropriately defined so that the function $\phi$ plays an analogous role to $\omega t$ in the simple harmonic oscillator.
%obeys the time dependent Schrödinger equation, with the Hamiltonian given by

%\begin{equation}
%\begin{split}
%\hat{H}=\frac{\hbar^2}{m}\left\{-\frac{1}{2}\partial_x^2-i \left(B+\frac{m}{\hbar}\dot{x}_0\right)\partial_x-\frac{i}{2}\mathcal{A}\left((x-x_0)\partial_x+\partial_x(x-x_0)\right)\right.\\
%\left.\frac{1}{2}\left[\left(A^R\right)^2+\frac{m}{\hbar}\dot{A}^I-\left(A^I\right)^2+2A^I\mathcal{A}\right]\left(x-x_0\right)^2+\left[B\mathcal{A}+\frac{m}{\hbar}\dot{B}\right](x-x_0)+\frac{1}{2}B^2\right\},
%\end{split}
%\end{equation}

At any given time instant, the wave-functions \eqref{eq:wavefunctions_general} resemble the usual Fock space states, i.e.,
\begin{equation}
\psi_n \left( x \right) = \frac{1}{\sqrt{2^n n!}} \left( \frac{m \omega^R}{\pi \hbar} \right)^{1/4} H_n \left( \sqrt{\frac{m \omega^R}{\hbar}} x \right) \exp \left[ - \frac{m \omega^R}{2 \hbar} x^2 \right]
\end{equation}
of the simple harmonic oscillator with eigenfrequency $\omega^R$, but they differ in the following three aspects:
\begin{itemize}
\item they have their argument shifted by $x_0$
\item they are multiplied by a position dependent phase,
\begin{equation}
\varphi \left( x \right) = - \frac{m \omega^I}{2 \hbar} \left( x - x_0 \right)^2 + \frac{p_0}{\hbar} \left( x - x_0 \right) ,
\end{equation}
which is \emph{identical for all states}
\item they are multiplied by \emph{a position independent phase}
\begin{equation}
\phi_n = - \left( n + \frac{1}{2} \right) \phi ,
\end{equation}
which depends on the state.
\end{itemize}
That is to say,
\begin{equation}
\left\langle x \middle| \Psi_n \right\rangle = \psi_n \left( x - x_0 \right) \exp \left[ i \varphi \left( x \right) \right] \exp \left( i \phi_n \right) .
\end{equation}
It follows that although these states are time-dependent, at any given time instant, they form an orthonormal basis for square integrable wavefunctions.

Orthonormality is a direct consequence of the orthonormality of the Fock space states,
\begin{equation}
\begin{split}
\left\langle \Psi_m \middle| \Psi_n \right\rangle &= \exp \left( i \phi_n - i \phi_m \right) \int dx \bar{\psi}_m \left( x - x_0 \right) \exp \left[ - i \varphi \left( x \right) \right] \psi_n \left( x - x_0 \right) \exp \left[ i \varphi \left( x \right) \right] \\
&= \exp \left( i \phi_n - i \phi_m \right) \int dx \bar{\psi}_m \left( x - x_0 \right) \psi_n \left( x - x_0 \right) \\
&= \exp \left( i \phi_n - i \phi_m \right) \int dx \bar{\psi}_m \left( x \right) \psi_n \left( x \right) \\
&= \exp \left( i \phi_n - i \phi_m \right) \delta_{mn} = \delta_{mn} .
\end{split}
\end{equation}

Regarding the completeness of the states, assume that we want to expand some square integrable function $f \left( x \right)$ as a linear combination of $\Psi_n$ at $t=t_0$. It is a well-known fact that the Fock space states $\psi_n \left( x \right)$ form a basis for square integrable wavefunctions. A shift in the argument of a wavefunction does not alter the fact that it is square integrable. The same holds for multiplication of a wavefunction with a pure phase. Therefore, we can always expand the wavefunction $f \left( x + x_0 \right) \exp \left[ - i \varphi \left( x + x_0 \right) \right]$ in the Fock space states,
\begin{equation}
f \left( x + x_0 \right) \exp \left[ - i \varphi \left( x + x_0 \right) \right] = \sum_{n = 0}^\infty c_n \psi_n \left( x \right) .
\end{equation}
This directly implies that
\begin{equation}
\begin{split}
f \left( x \right) &= \sum_{n = 0}^\infty c_n \psi_n \left( x - x_0 \right) \exp \left[ i \varphi \left( x \right) \right] \\
&= \sum_{n = 0}^\infty \left[ c_n \exp \left( - i \phi_n \right) \right] \psi_n \left( x - x_0 \right) \exp \left[ i \varphi \left( x \right) \right] \exp \left( i \phi_n \right) \\
&= \sum_{n = 0}^\infty \left[ c_n \exp \left( - i \phi_n \right) \right] \left\langle x \middle| \Psi_n \right\rangle ,
\end{split}
\end{equation}
i.e. any square integrable function can be expanded as a linear combination of the states $\left| \Psi_n \right\rangle$. Summing up, \emph{at any time instant, the states \eqref{eq:wavefunctions_general} form an orthonormal basis for square integrable wavefunctions}.

In the form of a bilinear kernel, the completeness of these wave-functions reads
\begin{equation}
\left\langle x \middle| \hat{I} \middle| x^\prime \right\rangle = \sum_{n=0}^\infty \left\langle x \middle| \Psi_n \right\rangle \left\langle \Psi_n \middle| x^\prime \right\rangle
\label{eq:completeness_kernel}
\end{equation}
and implies that
\begin{equation}
\left\langle x \middle| \hat{A}^\dagger \hat{A} \middle| x^\prime \right\rangle = \sum_{n=0}^\infty n \left\langle x \middle| \Psi_n \right\rangle \left\langle \Psi_n \middle| x^\prime \right\rangle .
\label{eq:AdagA_kernel}
\end{equation}

Before, we conclude this section, we would like to make an important comment. One can easily observe in the form of the wavefunctions \eqref{eq:wavefunctions_general} that they actually do not depend separately on $m$ and $\omega^R + i \omega^I$, but only on their product. Therefore, there is a whole family of Hamiltonians, one for each value of $m$, which have exactly the same solutions, and, thus, they are completely equivalent, as long as the time evolution of the system is concerned. These Hamiltonians may look quite different. Consider for example the case that $\omega^R$, $\omega^I$, $x_0$ and $p_0$ have been chosen so that $\mathcal{A} = 0$ and $\dot{x}_0 = p_0$. Then, if we write down the Hamiltonian with $m = 1$, there will be no linear term in $\hat{p}$. However, if we write it down with any other value of $m$, such a term will be present.

\section{A Pair of Harmonic Oscillators at a Coherent State}
\label{sec:two}
Entanglement entropy of coherent states was studied in the framework of quantum field theory in \cite{Benedict:1995yp} using the formalism of \cite{Callan:1994py}. It turns out that the entanglement entropy is identical to that in the case of the ground state. This is somehow expected, since the coherent states, including the ground state, are the classicalmost states of the system. These findings are in line to the analysis of the next two sections.

\subsection{The Reduced Density Matrix}
\label{subsec:two_density}
Before we proceed to study more complex systems, we first consider a system of two coupled harmonic oscillators. As subsystem $A$ we consider the harmonic oscillator, which is described by the coordinate $x_2$. The complementary system $A^C$ is consequently identical to the other oscillator, which is described by the coordinate $x_1$. The corresponding canonical momenta are named $p_2$ and $p_1$ respectively. Without loss of generality, the mass of both oscillators is considered to be equal to one. The Hamiltonian of this system is given by
\begin{equation}
H = \frac{1}{2} \left[ {p_1^2 + p_2^2 + {k_0} \left( {x_1^2 + x_2^2} \right) + {k_1} {\left( {{x_1} - {x_2}} \right)}^2} \right] .
\end{equation}
We define the canonical coordinates as usual
\begin{equation}
x_\pm := \frac{x_1 \pm x_2}{\sqrt{2}} , \quad p_\pm := \frac{p_1 \pm p_2}{\sqrt{2}} .
\label{eq:normal_coordinates}
\end{equation}
Then, the Hamiltonian assumes the form
\begin{equation}
H = \frac{1}{2}\left( {p_+^2 + p_-^2 + \omega_+^2 x_+^2 + \omega_-^2 x_-^2} \right) ,
\end{equation}
where $\omega_+ = \sqrt{k_0}$ and $\omega_- = \sqrt{k_0 + 2 k_1}$. This Hamiltonian describes two decoupled oscillators, one for each normal mode of the system.

We consider states of the composite system, where \emph{all} normal modes lie at coherent states. The wave function of a coherent state of an oscillator with eigenfrequency $\omega$ reads (see e.g. Appendix \ref{sec:squeezed})
%\begin{equation}
%\Psi=\left(\frac{\omega}{\pi}\right)^{1/4} \exp\left[-\left(\sqrt{\frac{\omega}{2}}x-\lambda\right)^2+\frac{1}{2}\left(\lambda^2-\vert\lambda\vert^2\right)-\frac{i}{2}\omega t\right],
%\end{equation}
%where $\lambda=\lambda_0 \exp\left(-i \omega t\right).$ The phase $\frac{i}{2}\omega t$ is included for completeness, since this is the form of wavefunction that solves the time-dependent Schrödinger equation. Yet this phase is unphysical and will cancel at the level of the density matrix, thus we drop it in anything that follows. It would be convenient to rescale $\lambda$ as
%\begin{equation}
%\lambda\rightarrow \sqrt{\frac{\omega}{2}}\lambda
%\end{equation}
%to yield
\begin{equation}
\left\langle x \middle| \Psi \left( \lambda \right) \right\rangle = \left( \frac{\omega}{\pi} \right)^{1/4} \exp \left[ - \frac{\omega}{2} \left( x - \lambda \right)^2 + \frac{\omega}{4} \left( \lambda^2 - \left| \lambda \right|^2 \right) - i \frac{\omega}{2} t \right] .
\end{equation}
The parameter $\lambda$ evolves with time as
\begin{equation}
\lambda = \lambda_0 e^{- i \omega t}
\end{equation}
and determines the mean values of position and momentum
\begin{equation}
\left\langle x \right\rangle = \re \lambda , \quad \left\langle p \right\rangle = \omega \im \lambda .
\end{equation}

It directly follows that the wavefunction of the composite system when both normal modes lie at a coherent state reads
\begin{multline}
\left\langle x_+ , x_- \middle| \Psi \left( \lambda_+ , \lambda_- \right) \right\rangle \\
= \left( \frac{\omega_+ \omega_-}{\pi^2} \right)^{1/4} \exp\left[ - \frac{\omega_+}{2} \left( x_+ - \lambda_+ \right)^2 + \frac{\omega_+}{4}\left( \lambda_+^2 - \left| \lambda_+ \right|^2 \right) - i \frac{\omega_+}{2} t\right]\\
\times \exp \left[ - \frac{\omega_-}{2} \left( x_- - \lambda_- \right)^2 + \frac{\omega_-}{4} \left( \lambda_-^2 - \left| \lambda_- \right|^2 \right) - i\frac{\omega_-}{2} t \right] .
\label{eq:Psi_system_coh}
\end{multline}
The density matrix of the composite system reads
\begin{multline}
\left\langle x_+ , x_- \middle| \hat{\rho} \left( \lambda_+ , \lambda_- \right) \middle| x_+^\prime , x_-^\prime \right\rangle \\
= \left( \frac{\omega_+ \omega_-}{\pi^2} \right)^{1/2} \exp \left[ - \frac{\omega_+}{2} \left( x_+ - \lambda_+ \right)^2 - \frac{\omega_+}{2} \left( x^\prime_+ - \bar{\lambda}_+ \right)^2 + \frac{\omega_+}{4} \left(\lambda_+ - \bar{\lambda}_+ \right)^2 \right] \\
\times \exp \left[ - \frac{\omega_-}{2} \left( x_- - \lambda_- \right)^2 - \frac{\omega_-}{2} \left( x^\prime_- - \bar{\lambda}_- \right)^2 + \frac{\omega_-}{4}\left( \lambda_- - \bar{\lambda}_- \right)^2 \right] .
\end{multline}
In order to obtain the reduced density matrix of the second oscillator, which comprises the subsystem $A$, we have to trace out the degree of freedom corresponding to the first harmonic oscillator,
\begin{equation}
\left\langle x_2 \middle| \hat{\rho}_2 \middle| x_2^\prime \right\rangle = \int dx_1 \left\langle x_1 , x_2 \middle| \hat{\rho} \middle| x_1 , x_2^\prime \right\rangle .
\end{equation}
This is a simple task of performing a Gaussian integral, but first we need to revert to the original coordinates $x_1$ and $x_2$ from the normal mode coordinates $x_\pm$. It is also convenient to define the moduli $\lambda_1$ and $\lambda_2$ so that
\begin{equation}
\lambda_\pm = \frac{1}{\sqrt{2}} \left( \lambda_1 \pm \lambda_2 \right) ,
\end{equation}
in direct analogy to the definition of the normal mode coordinates \eqref{eq:normal_coordinates}. Bear in mind that the ``local'' coherent state parameters $\lambda_1$ and $\lambda_2$ do not have the simple time evolution of those of the normal modes. The reduced density matrix reads
\begin{equation}
\left\langle x_2 \middle| \hat{\rho}_2 \left( \lambda_1 , \lambda_2 \right) \middle| x_2^\prime \right\rangle = \left( \frac{\gamma - \beta}{\pi} \right)^{1/2} \exp \left[ - \frac{\gamma}{2} \left( y_2^2 + y^{\prime 2}_2 \right) + \beta y_2 y^\prime_2 + i \delta \left( y_2 - y^\prime_2 \right) \right] ,
\label{eq:reduced_rho_pair_coh}
\end{equation}
where we used the arguments $\lambda_1$ and $\lambda_2$ to indicate that this is the reduced density matrix that corresponds to the coherent state \eqref{eq:Psi_system_coh} and
\begin{align}
y_2 &= x_2 - \re \lambda_2 , \qquad y^\prime_2 = x^\prime_2 -\re \lambda_2 , \\
\beta &= \frac{\left( \omega_+ - \omega_- \right)^2}{4 \left( \omega_+ +\omega_- \right)} , \\
\gamma &= \frac{\omega_+^2 + 6 \omega_+ \omega_- + \omega_-^2}{4 \left( \omega_+ + \omega_- \right)} , \\
\delta &= \frac{1}{2} \left[ \left( \omega_+ - \omega_- \right) \im \lambda_1 + \left( \omega_+ + \omega_- \right) \im \lambda_2 \right] .
\end{align}

The spectrum of a density matrix $\hat{\rho}$ is in principle completely determined by the sequence of the traces $\tr \left( \hat{\rho}^n \right)$ for all $n \in \mathbb{N}$.
As the parameters $\beta$ and $\gamma$ \emph{do not depend} on the values of values of $\lambda_1$ and $\lambda_2$, it is evident that
\begin{equation}
\tr \left[ \hat{\rho}_2^n \left( \lambda_1 , \lambda_2 \right) \right] = \tr \left[ \hat{\rho}_2^n \left( 0 , 0 \right) \right] , \quad \forall n \in \mathbb{N} .
\end{equation}
Thus, the spectrum of the reduced density matrix coincides with the spectrum of the reduced density matrix when the overall system lies at its ground state. It follows that the entanglement entropy is also identical to the entanglement entropy in the case of the ground state. Nevertheless, this neither implies that the former reduced density matrix coincides with the latter nor that their dynamics coincide.

We can verify the above and obtain the spectrum of the reduced density matrix in a more direct way, which also provides its eigenfunctions. Mehler's formula for Hermite polynomials states that
\begin{multline}
\sum_{n=0}^\infty \frac{1}{n!} \left( \frac{\xi}{2} \right)^n H_n \left( x \right) H_n \left( y \right) \exp \left[ - \frac{1}{2} \left( x^2 + y^2 \right) \right] \\
= \frac{1}{\sqrt{1 - \xi^2}} \exp \left[ - \frac{\left( 1 + \xi^2 \right) \left( x^2 + y^2 \right) - 4 \xi x y}{2 \left( 1 - \xi^2 \right)} \right] .
\label{eq:Mehler}
\end{multline}
So identifying
\begin{equation}
\xi = \frac{\beta}{\gamma + \alpha} , \qquad x = \sqrt{\alpha} y_2 , \qquad y = \sqrt{\alpha} y_2^\prime ,
\end{equation}
where we have defined
\begin{equation}
\alpha := \sqrt{\gamma^2 - \beta^2} = \sqrt{\omega_+ \omega_-} ,
\end{equation}
we may directly apply Mehler's formula to the reduced density matrix \eqref{eq:reduced_rho_pair_coh} in order to re-express it as
\begin{multline}
\left\langle x_2 \middle| \hat{\rho}_2 \middle| x_2^\prime \right\rangle = \sqrt{\frac{\alpha}{\pi}} \sum_{n=0}^\infty \frac{1 - \xi}{n!}\left( \frac{\xi}{2} \right)^n H_n \left( \sqrt{\alpha} y_2 \right) H_n \left( \sqrt{\alpha} y^\prime_2 \right) \\
\times \exp \left[ - \frac{\alpha}{2} \left( y_2^2 + y^{\prime 2}_2 \right) + i \delta \left( y_2 - y^\prime_2 \right) \right] .
\end{multline}
As a result
\begin{equation}
\left\langle x_2 \middle| \hat{\rho}_2 \middle| x_2^\prime \right\rangle = \left( 1 - \xi \right) \sum_{n=0}^\infty \xi^n \psi_n \left( x_2 - \re \lambda_2 \right) \psi_n \left( x^\prime_2 - \re \lambda_2 \right) \exp \left[ i \delta \left( x_2 - x^\prime_2 \right) \right] ,
\end{equation}
where $\psi_n$ are the energy eigenstates of an effective simple harmonic oscillator with eigenfrequency equal to $\omega_{\text{eff}} = \alpha$. We may express the reduced density matrix as 
\begin{equation}\label{eq:reduced_two_eigenstates}
\left\langle x_2 \middle| \hat{\rho}_2 \middle| x_2^\prime \right\rangle = \left( 1 - \xi \right) \sum_{n=0}^\infty \xi^n \Psi_n \left( x_2 \right) \bar{\Psi}_n \left( x^\prime_2 \right) ,
\end{equation}
where
\begin{equation}\label{eq:state}
\Psi_n \left( x \right) = \psi_n \left( x - \re \lambda_2 \right) \exp \left[ i \delta \left( x - \re \lambda_2 \right) - i \alpha t \left( n + \frac{1}{2} \right) \right] .
\end{equation}
The orthonormality of the eigenstates of the harmonic oscillator, combined with the fact that the phase $\delta$ is \emph{the same} for all states $\Psi_n$ implies that the states $\Psi_n$ are also orthonormal. It follows that equation \eqref{eq:reduced_two_eigenstates} provides the expansion of the reduced density matrix into its eigenfunctions.

It is obvious that the eigenvalues of the reduced density matrix
\begin{equation}\label{eq:eigs_pair}
p_n = \left( 1 - \xi \right) \xi^n
\end{equation}
are independent of $\lambda_\pm$, implying that they are time-independent and equal to the eigenvalues of the reduced system when the composite system lies in the vacuum state. 

\subsection{Time Evolution of the Reduced System}
\label{subsec:two_effective}
The eigenfunctions \eqref{eq:state} of the reduced density matrix \eqref{eq:reduced_rho_pair_coh} belong to the family of wave-functions \eqref{eq:wavefunctions_general} that solve the generalized oscillator system, which we presented in Section \ref{sec:Osc}. In particular they are identical under the identification
\begin{equation}
m \omega^R = \alpha ,\quad \omega^I = 0 , \quad x_0 = \re \lambda_2 , \quad p_0 = \delta ,
\label{eq:identification_two}
\end{equation}
where $\hbar$ is set to one.
Therefore, all the eigenfunctions \eqref{eq:state} are solutions of the same time-dependent Schrödinger equation, where the corresponding Hamiltonian is given by \eqref{eq:H_x_p}, i.e.
\begin{multline}
\hat{H}_2 = \frac{1}{2m} \left( \hat{p} - \delta + m \re \dot{\lambda}_2 \right)^2 + \frac{1}{2 m} \alpha^2 \left( \hat{x} - \re \lambda_2 \right)^2 \\
- \dot{\delta} \left( \hat{x} - \re \lambda_2 \right) + \re \dot{\lambda}_2 \left( \delta - \frac{1}{2} m \re \dot{\lambda}_2 \right) .
\label{eq:H_eff_2_coh_pre}
\end{multline}

It is interesting to find the time derivative of the function $x_0 \left( t \right)$. Bearing in mind that the coherent state parameters for the normal modes have a very simple time dependence, namely $\lambda_\pm \left( t \right) = \lambda_\pm \left( 0 \right) e^{- i \omega_\pm t} $, it is not difficult to show that
\begin{equation}
\begin{split}
\dot{x}_0 &= \re \dot{\lambda}_2 = \frac{1}{\sqrt{2}} \re \left( \dot{\lambda}_+ - \dot{\lambda}_- \right) \\
&= \frac{1}{\sqrt{2}} \re \left( - i \omega_+ \lambda_+ + i\omega_- \lambda_- \right) \\
&= \frac{1}{2} \left[ \omega_+ \im \left( \lambda_1 + \lambda_2 \right) - \omega_- \im \left( \lambda_1 - \lambda_2 \right) \right] = \delta = p_0 ,
\end{split}
\end{equation}
This implies that the Hamiltonian \eqref{eq:H_eff_2_coh_pre} is simplified if we chose $m = 1$. Then, it reads
\begin{equation}
\hat{H}_2 = \frac{1}{2} \hat{p}_2^2 + \frac{1}{2} \alpha^2 \left( \hat{x}_2 - \re \lambda_2 - \frac{\dot{\delta}}{\alpha^2} \right)^2 + \frac{1}{2} \delta^2 - \frac{1}{2} \frac{\dot{\delta}^2}{\alpha^2} .
\label{eq:H_eff_2_coh}
\end{equation}

After tracing out the first harmonic oscillator, the second one comprises an open quantum system. The time evolution of such a system is non-unitary in general. Nevertheless, since \emph{all} the eigenstates of the reduced density matrix obey the equation
\begin{equation}
i\hbar\dot{\Psi}_n = \hat{H}_2 \Psi_n ,
\end{equation}
where $\hat{H}_2$ is given by \eqref{eq:H_eff_2_coh}, equation \eqref{eq:reduced_two_eigenstates} directly implies that
\begin{equation}
i\hbar\dot{\hat{\rho}}_2=\left[\hat{H}_2,\hat{\rho}_2\right],
\end{equation}
meaning that when the overall system lies in a coherent state, the time evolution of the reduced system \emph{is unitary}.

How can we conceive this effective Hamiltonian at a classical level? This Hamiltonian describes a force from an effective spring with time-independent spring constant equal to $a^2$, but time-dependent equilibrium position which is equal to $x_0 + \dot{p}_0 / a^2$, i.e. the spring force reads
\begin{equation}
F_{\mathrm{eff}} = - a^2 \left(x_2 - x_0 \right) - \dot{p}_0 .
\end{equation}
If we associate the coherent state to a classical motion, so that the classical values of the coordinates are given by their mean values of their quantum counterparts, which are described by the coherent state, (this is well-known to be indeed a solution of the classical equations of motion), then, this force will simulate exactly the force from both real springs that exert forces on the second oscillator, i.e.
\begin{equation}
F_{\mathrm{eff}} = - k_0 x_2 - k_1 \left(x_2 - x_1\right) = F_2 ,
\end{equation}
including the one which connects it to the invisible first oscillator.

So classically this effective Hamiltonian is a nice one-particle model that recovers the correct motion of the second observable oscillator as part of the composite system, accumulating the effect from the first invisible oscillator in the time-dependence of the equilibrium position of the single effective spring. Although this effective Hamiltonian \emph{does depend on the state of the overall system}, the interesting observation is that this dependence appears solely in the effective equilibrium position of the effective spring and not at all at its constant. No matter what is the coherent state that we prepare the composite system, the constant of the effective spring is the same; only the specific time-dependence of its equilibrium position is altered.

\subsection{The Modular Hamiltonian}
\label{subsec:two_modular}
We recall that the modular Hamiltonian is defined as the logarithm of the reduced density matrix
\begin{equation}
\hat{\rho}=e^{- \hat{H}_M}.
\end{equation}
So employing equation \eqref{eq:reduced_two_eigenstates}, we obtain
\begin{equation}
\left\langle x^\prime_2 \middle| H_M \middle| x_2 \right\rangle = - \sum_{n=1}^\infty \left( n \ln \xi + \ln \left( 1 - \xi \right) \right) \Psi_n \left( x_2 \right) \bar{\Psi}_n \left( x^\prime_2 \right) .
\end{equation}
As a direct consequence of the relations \eqref{eq:completeness_kernel} and \eqref{eq:AdagA_kernel} the modular Hamiltonian assumes the form
\begin{equation}
\hat{H}_M = - \ln \xi \hat{A}^\dagger \hat{A} - \ln \left( 1 - \xi \right) \hat{I} ,
\end{equation}
where the parameters in the generalized creation and annihilation operators are given by equation \eqref{eq:identification_two}.

Employing equation \eqref{eq:AdagA_x_p}, we may write the modular Hamiltonian in terms of the position and momentum operators as,
\begin{equation}
\hat{H}_M = - \ln \xi \left( \frac{1}{2 m \alpha} \left( \hat{p}_2 - p_0 \right)^2 +\frac{\alpha}{2 m} \left( \hat{x}_2 - x_0 \right)^2 \right) + \frac{1}{2} \ln \xi - \ln \left( 1 - \xi \right) .
\end{equation}

This Hamiltonian, when seen as a classical Hamiltonian, reproduces the correct time evolution of the classical system, like the effective Hamiltonian that we presented in Section \ref{subsec:two_effective}, only when $\ln \xi = \alpha$. However, we should not expect this to happen. The physical meaning of the modular Hamiltonian is rather thermodynamic and assumes form through the first law of entanglement thermodynamics.

\section{System of Harmonic Oscillators at Coherent State}
\label{sec:many}
In this section, we treat an arbitrary system of coupled oscillators. The analysis follows closely the original presentation of \cite{srednicki}. 
\subsection{The Reduced Density Matrix}
Assume a system of $N$ coupled harmonic oscillators described by the quadratic Hamiltonian
\begin{equation}
H = \frac{1}{2} \mathbf{p}^T \mathbf{p} + \frac{1}{2} \mathbf{x}^T K \mathbf{x} ,
\end{equation}
where $\mathbf{x}$ and $\mathbf{p}$ stand for $N$-dimensional column matrices containing the coordinates and momenta of the $N$ harmonic oscillators. The matrix $K$ is symmetric and has positive eigenvalues, so that the system is stable. As $K$ is positively defined, we can define the matrix $\Omega$ as the square root of $K$, which has all its eigenvalues positive.

In the following, we consider as subsystem $A$ the $N-n$ oscillators, which are described by the coordinates $x_i$ with $i > n$. The complementary subsystem $A^C$ comprises of the $n$ oscillators, which are described by the coordinates $x_i$ with $i \le n$.

Following the usual normal mode analysis, there is an orthogonal transformation $O$, relating the coordinates $\mathbf{x}$ to the normal coordinates $\tilde{\mathbf{x}}$, which diagonalizes the matrix $K$, reducing the system to a set of decoupled harmonic oscillators, one for each normal mode. In other words,
\begin{equation}
\tilde{\mathbf{x}} = O \mathbf{x},\quad \tilde{\mathbf{p}} = O \mathbf{p} ,
\end{equation}
such that
\begin{equation}
H =\frac{1}{2} \tilde{\mathbf{p}}^T \tilde{\mathbf{p}}  + \frac{1}{2} \tilde{\mathbf{x}}^T K^D \tilde{\mathbf{x}} .
\end{equation}
The diagonalized matrix $K^D$
\begin{equation}
{K^D_{ij}} \equiv {\omega _i}^2{\delta _{ij}},
\end{equation}
is related to the initial matrix $K$ as,
\begin{equation}
K = {O^T}{K^D}O .
\end{equation}

As in Section \ref{sec:two}, we want to study states of the overall system, where \emph{all} normal modes lie in a coherent state. Such a state reads
\begin{multline}
\left\langle \mathbf{x} \middle| \Psi \left( \tilde{\boldsymbol{\lambda}} \right) \right\rangle = \sqrt[4]{\frac{\det \Omega_D}{\pi^N}} \exp \left[ - \frac{1}{2} \left( \tilde{\mathbf{x}} - \tilde{\boldsymbol{\lambda}} \right)^T \Omega^D \left( \tilde{\mathbf{x}} - \tilde{\boldsymbol{\lambda}} \right) \right. \\
\left. + \frac{1}{4} \left( \tilde{\boldsymbol{\lambda}}^T \Omega^D \tilde{\boldsymbol{\lambda}} - \tilde{\boldsymbol{\lambda}}^\dagger \Omega^D \tilde{\boldsymbol{\lambda}} \right) - \frac{i}{2} \tr \Omega^D t \right] ,
\end{multline}
where $\Omega^D_{ij} = \omega_i \delta_{ij}$, is the square root of $K^D$. In an obvious manner, it is connected to the matrix $\Omega$ as
\begin{equation}
\Omega = O^T \Omega^D O .
\end{equation}
The column matrix $\tilde{\boldsymbol{\lambda}}$ contains the complex ``canonical'' moduli of the coherent states of the $N$ normal modes. It evolves with time as
\begin{equation}
\tilde{\boldsymbol{\lambda}} \left( t \right) = e^{- i \Omega^D t} \tilde{\boldsymbol{\lambda}} \left( 0 \right) .
\label{eq:lambda_evolution}
\end{equation}
We define the moduli $\boldsymbol{\lambda}$, so that they are connected to the ``canonical'' moduli as
\begin{equation}
\boldsymbol{\lambda} = O^T \tilde{\boldsymbol{\lambda}} ,
\end{equation}
i.e. in exactly the same way as the original coordinates are related to the canonical ones. We rewrite the wavefunction in terms of the original coordinates and the moduli $\boldsymbol{\lambda}$ to get
\begin{equation}
\left\langle \mathbf{x} \middle| \Psi \left( \boldsymbol{\lambda} \right) \right\rangle =  \sqrt[4]{\frac{\det \Omega}{\pi^N}} \exp \left[ - \frac{1}{2} \left( \mathbf{x} - \boldsymbol{\lambda} \right)^T \Omega \left( \mathbf{x} - \boldsymbol{\lambda} \right) + \frac{1}{4} \left( \boldsymbol{\lambda}^T \Omega \boldsymbol{\lambda} - \boldsymbol{\lambda}^\dagger \Omega \boldsymbol{\lambda} \right) - \frac{i}{2} \tr \Omega t \right] .
\end{equation}
The density matrix reads
\begin{multline}
\left\langle \mathbf{x} \middle| \rho \left( \boldsymbol{\lambda} \right) \middle| \mathbf{x}^\prime \right\rangle = \sqrt{\frac{\det \Omega}{\pi^N}} \exp \left[ - \frac{1}{2} \left( \mathbf{x} - \boldsymbol{\lambda} \right)^T \Omega \left( \mathbf{x} - \boldsymbol{\lambda} \right) - \frac{1}{2} \left( \mathbf{x}^{\prime} - \bar{\boldsymbol{\lambda}} \right)^T \Omega \left( \mathbf{x}^\prime - \bar{\boldsymbol{\lambda}} \right) \right] \\
\times \exp \left[ \frac{1}{4} \left( \boldsymbol{\lambda} - \bar{\boldsymbol{\lambda}} \right)^T \Omega \left( \boldsymbol{\lambda} - \bar{\boldsymbol{\lambda}} \right) \right] .
\end{multline}

In the following, we use the block notation
\begin{equation}
\label{eq:block_notation}
\mathbf{x}=\begin{pmatrix}
\mathbf{x}_C\\
\mathbf{x}_A\\
\end{pmatrix} , \quad \boldsymbol{\lambda}=\begin{pmatrix}
\boldsymbol{\lambda}_C\\
\boldsymbol{\lambda}_A\\
\end{pmatrix} , \quad \Omega= \begin{pmatrix}
\Omega_C & \Omega_B\\
\Omega_B^T & \Omega_A\\
\end{pmatrix} .
\end{equation}
The matrix columns $\mathbf{x}_C$ and $\boldsymbol{\lambda}_C$ are $n$-dimensional and contain the coordinates and corresponding coherent state moduli of the degrees of freedom of subsystem $A^C$. Similarly, the matrix columns $\mathbf{x}_A$ and $\boldsymbol{\lambda}_A$ are $\left(N-n\right)$-dimensional and contain the coordinates and corresponding coherent state moduli of the degrees of freedom of subsystem $A$. Finally, in an obvious manner, the blocks of the matrix $\Omega$ have been defined so that they have the appropriate dimensions: $\Omega_C$ is an $n\times n$ matrix, $\Omega_B$ is $n\times \left( N - n \right)$ and $\Omega_A$ is $\left( N - n \right)\times \left( N - n \right)$.

Once we have introduced this block formalism, we can proceed to find the reduced density matrix of subsystem $A$. This reads
\begin{equation}
\left\langle \mathbf{x}_A \middle| \rho_A \middle| \mathbf{x}^\prime_A \right\rangle = \int dx_C \left\langle \mathbf{x}_A , \mathbf{x}_C \middle| \rho \middle| \mathbf{x}^\prime_A , \mathbf{x}_C \right\rangle .
\end{equation}
It is a matter of tedious but straightforward algebra to show that this yields
\begin{multline}
\left\langle \mathbf{x}_A \middle| \rho_A \left( \boldsymbol{\lambda} \right) \middle| \mathbf{x}^\prime_A \right\rangle = \sqrt{\frac{\det \Omega}{\pi^N}} \exp \left[ - \frac{1}{2} \left( \mathbf{x}_A - \mathbf{x}_{0A} \right)^T \gamma \left( \mathbf{x}_A - \mathbf{x}_{0A} \right) \right. \\
\left. - \frac{1}{2} \left( \mathbf{x}^\prime_A - \mathbf{x}_{0A} \right)^T \gamma \left( \mathbf{x}^\prime_A - \mathbf{x}_{0A} \right) + \left( \mathbf{x}^\prime_A - \mathbf{x}_{0A} \right)^T \beta \left( \mathbf{x}_A - \mathbf{x}_{0A} \right) + i \left( \mathbf{x}_A - \mathbf{x}^\prime_A \right)^T \boldsymbol{\delta} \right] \\
\times \int d \mathbf{x}_C \exp \left[ - \left( \mathbf{x}_C - \mathbf{x}_{0C} \right)^T \Omega_C \left( \mathbf{x}_C - \mathbf{x}_{0C} \right) \right] ,
\end{multline}
where
\begin{align}
\mathbf{x}_{0C} &= \re \boldsymbol{\lambda}_C + \Omega_C^{-1} \Omega_B \left( \re \boldsymbol{\lambda}_A - \frac{\mathbf{x}_A + \mathbf{x}^\prime_A}{2} \right) , \\
\mathbf{x}_{0A} &= \re \boldsymbol{\lambda}_A , \\
\beta& = \frac{1}{2} \Omega_B^T \Omega_C^{-1} \Omega_B , \\
\gamma &= \Omega_A - \beta ,\\
\boldsymbol{\delta} &= \Omega_B^T \im \boldsymbol{\lambda}_C + \Omega_A \im \boldsymbol{\lambda}_A .
\end{align}
Notice that the matrices $\gamma$ and $\beta$ are independent of the moduli $\boldsymbol{\lambda}$. Therefore, they do not depend on the coherent state of the overall system; they are the same for all coherent states.

The Gaussian integral can trivially be performed to obtain
\begin{multline}
\left\langle \mathbf{x}_A \middle| \rho_A \left( \boldsymbol{\lambda} \right) \middle| \mathbf{x}^\prime_A \right\rangle = \sqrt{\frac{\det \left( \gamma - \beta \right)}{\pi^{\left( N - n \right)}}} \exp \left[ - \frac{1}{2} \left( \mathbf{x}_A - \mathbf{x}_{0A} \right)^T \gamma \left( \mathbf{x}_A - \mathbf{x}_{0A} \right) \right. \\
\left. - \frac{1}{2} \left( \mathbf{x}^\prime_A - \mathbf{x}_{0A} \right)^T \gamma \left( \mathbf{x}^\prime_A - \mathbf{x}_{0A} \right) + \left( \mathbf{x}^\prime_A - \mathbf{x}_{0A} \right)^T \beta \left( \mathbf{x}_A - \mathbf{x}_{0A} \right) + i \left( \mathbf{x}_A - \mathbf{x}^\prime_A \right)^T \boldsymbol{\delta} \right] .
\end{multline}

The matrices $\gamma$ and $\beta$ are by construction real symmetric $\left( N - n \right) \times \left( N - n \right)$ matrices. Let $\mathbf{v}_i$, $i = 1 , 2 , \ldots , N - n$, be a set of orthonormal eigenvectors of the symmetric matrix 
\begin{equation}
B := \gamma^{-\frac{1}{2}} \beta \gamma^{-\frac{1}{2}} ,
\end{equation}
i.e.
\begin{equation}
I_{N - n} = \sum_{i = 1}^{N - n} \mathbf{v}_i \mathbf{v}_i^T , \quad B = \sum_{i = 1}^{N - n} \beta^D_i \mathbf{v}_i \mathbf{v}_i^T ,
\end{equation}
where $\beta^D_i$ is the eigenvalue corresponding to the eigenvector $\mathbf{v}_i$.

We define
\begin{equation}
\tilde{x}_{Ai} := \mathbf{v}_i^T \gamma^{\frac{1}{2}} \mathbf{x}_A , \quad \tilde{x}^\prime_{Ai} := \mathbf{v}_i^T \gamma^{\frac{1}{2}} \mathbf{x}^\prime_A, \quad \tilde{x}_{0Ai} := \mathbf{v}_i^T \gamma^{\frac{1}{2}} \mathbf{x}_{0A} ,
\end{equation}
\begin{equation}
\tilde{y}_{Ai} := \tilde{x}_{Ai} - \tilde{x}_{0Ai}, \quad \tilde{y}^\prime_{Ai} := \tilde{x}^\prime_{Ai} - \tilde{x}_{0Ai}
\end{equation}
and finally
\begin{equation}
w_i := \mathbf{v}_i^T \gamma^{-\frac{1}{2}} \boldsymbol{\delta} .
\end{equation}
Using the above definitions, the reduced density matrix assumes the form
\begin{multline}
\left\langle \mathbf{x}_A \middle| \rho_A \left( \boldsymbol{\lambda} \right) \middle| \mathbf{x}^\prime_A \right\rangle \\
= \sqrt{\frac{\det \left( I - \beta^D \right)}{\pi^{\left( N - n \right)}}} \exp \left[ \sum_{i = 1}^{N - n} \left( - \frac{1}{2} \tilde{y}_{Ai}^2 - \frac{1}{2} \tilde{y}^{\prime 2}_{Ai} + \beta^D_i \tilde{y}_{Ai} \tilde{y}^\prime_{Ai} + i \left( \tilde{y}_{Ai} - \tilde{y}^\prime_{Ai} \right) w_i \right) \right] .
\end{multline}
We observe that the density matrix has assumed a form, where it can be written as the tensor product of $N - n$ matrices; each of these matrices is a function of only one of the coordinates $\tilde{x}_{Ai}$. For this reason we name the coordinates $\tilde{x}_{Ai}$ as ``the normal coordinates of the reduced system''. Having written down the reduced density matrix in this form, we may take advantage of Mehler's formula \eqref{eq:Mehler} and write the reduced density matrix as
\begin{multline}
\left\langle \mathbf{x}_A \middle| \rho_A \left( \boldsymbol{\lambda} \right) \middle| \mathbf{x}^\prime_A \right\rangle = \prod_{i = 1}^{N - n} \left( \frac{\alpha_i}{\pi} \right)^{1/2} \sum_{n_i = 0}^\infty \frac{1 - \xi_i}{n!} \left( \frac{\xi_i}{2} \right)^n H_n \left( \sqrt{\alpha_i} \tilde{y}_{Ai} \right) H_n \left( \sqrt{\alpha_i} \tilde{y}^\prime_{Ai} \right) \\
\times \exp \left[ - \frac{\alpha_i}{2} \left( \tilde{y}_{Ai}^2 + \tilde{y}^{\prime 2}_{Ai} \right) + i w_i \left( \tilde{y}_{Ai} - \tilde{y}^\prime_{Ai} \right) \right] .
\end{multline}
where we have defined
\begin{equation}
\alpha_i := \sqrt{1 - \left( \beta^D_i \right)^2}, \quad \xi_i = \frac{\beta^D_i}{1 + \alpha_i} .
\label{eq:a_i_xi_i}
\end{equation}
In other words,
\begin{equation}
\left\langle \mathbf{x}_A \middle| \rho_A \left( \boldsymbol{\lambda} \right) \middle| \mathbf{x}^\prime_A \right\rangle = \prod_{i = 1}^{N - n} \left( 1 - \xi_i \right)\sum_{n_i = 0}^\infty \xi_i^{n_i} \psi_{n_i} \left( \tilde{y}_{Ai} \right) \psi_{n_i} \left( \tilde{y}^\prime_{Ai} \right) \exp \left[ i w_i \left( \tilde{y}_{Ai} - \tilde{y}^\prime_{Ai} \right) \right] ,
\label{eq:reduced_rho_many_coh}
\end{equation}
where $\psi_{n_i}$ are the eigenfunctions of the Hamiltonian of an effective simple harmonic oscillator with eigenfrequency equal to $\omega_{\text{eff}} = \alpha_i$. We may further write the reduced density matrix as 
\begin{equation}
\left\langle \mathbf{x}_A \middle| \rho_A \left( \boldsymbol{\lambda} \right) \middle| \mathbf{x}^\prime_A \right\rangle = \prod_{i = 1}^{N - n} \left(1 - \xi_i \right) \sum_{n_i = 0}^\infty \xi_i^{n_i} \Psi_{n_i} \left( \tilde{y}_{Ai} \right) \bar{\Psi}_{n_i} \left( \tilde{y}^\prime_{Ai} \right) ,
\label{eq:reduced_many_eigenstates}
\end{equation}
where
\begin{equation}
\Psi_{n_i} \left( x \right) = \psi_{n_i} \left( x \right) \exp \left[ i w_i x-i \alpha_i t \left( n + \frac{1}{2} \right) \right] .
\label{eq:state_many}
\end{equation}
The orthonormality of the eigenstates of the harmonic oscillator, combined with the fact that the phase $w_i$ is \emph{the same} for all states $\Psi_{n_i}$ implies that the states $\Psi_{n_i}$ are also orthonormal. It follows that equation \eqref{eq:reduced_many_eigenstates} provides the expansion of the reduced density matrix into its eigenfunctions, which read
\begin{equation}
\Psi_{\left\{ n_i \right\}} = \prod_{i = 1}^{N - n} \Psi_{n_i} .
\end{equation}

It is obvious that the corresponding eigenvalues
\begin{equation}
p_{\left\{ n_i \right\}} = \prod_{i = 1}^{N - n} \left( 1 - \xi_i \right) \xi_i^{n_i}
\label{eq:eigs_many}
\end{equation}
are independent of $\boldsymbol{\lambda}$, implying that they are time-independent and equal to the eigenvalues of the reduced system when the composite system lies in the vacuum state. 

\subsection{Time Evolution of the Reduced System}
The eigenfunctions \eqref{eq:state} of the reduced density matrix \eqref{eq:reduced_rho_many_coh} belong to the family of wave-functions \eqref{eq:wavefunctions_general} that solve the generalized oscillator system that we presented in Section \ref{sec:Osc}. In particular they coincide under the identification
\begin{equation}
m_i \omega_i^R = \alpha_i ,\quad \omega_i^I = 0 ,\quad \tilde{x}_{0Ai} = \re \left( \mathbf{v}_i^T \gamma^{\frac{1}{2}} \boldsymbol{\lambda}_{A} \right) , \quad \tilde{p}_{0Ai} = w_i = \mathbf{v}_i^T \gamma^{-\frac{1}{2}} \boldsymbol{\delta} ,
\label{eq:identification_many}
\end{equation}
where $\hbar$ is set to one. It follows that the corresponding Hamiltonian is
\begin{multline}
\hat{H}_A = \sum_{i = 1}^{N - n} \frac{1}{2 m_i} \left( \hat{\tilde{p}}_{Ai} - w_i + m_i \dot{\tilde{x}}_{0Ai} \right)^2 + \frac{1}{2 m_i} \alpha_i^2 \left( \hat{\tilde{x}}_{Ai} - \tilde{x}_{0Ai} \right)^2 \\
- \dot{w}_i \left( \hat{\tilde{x}}_{Ai} - \tilde{x}_{0Ai} \right) + \dot{\tilde{x}}_{0Ai} \left( w_i - \frac{1}{2} m_i \dot{\tilde{x}}_{0Ai} \right) .
\label{eq:Heffective_modes}
\end{multline}
We have the freedom of determining freely the value of each $m_i$ independently. 

As a direct consequence of \eqref{eq:lambda_evolution}, we get
\begin{equation*}
\re \dot{\tilde{\boldsymbol{\lambda}}} = \Omega^D \im \tilde{\boldsymbol{\lambda}} .
\end{equation*}
This implies that
\begin{equation*}
\re O^T \dot{\tilde{\boldsymbol{\lambda}}} = O^T \Omega^D O \im O^T \tilde{\boldsymbol{\lambda}} ,
\end{equation*}
i.e.
\begin{equation*}
\re \dot{\boldsymbol{\lambda}} = \Omega \im \boldsymbol{\lambda} .
\end{equation*}
Employing the block form notation \eqref{eq:block_notation} yields
\begin{equation*}
\re \dot{\boldsymbol{\lambda}}_A = \Omega_B^T \im \boldsymbol{\lambda}_C + \Omega_C \im \boldsymbol{\lambda}_A = \boldsymbol{\delta} .
\end{equation*}

Finally, we need to revert back to the original coordinates $\mathbf{x}_{0A}$. It is a matter of tedious algebra to show that
%\begin{multline}
%\hat{H}_A = \frac{1}{2} \left( \mathbf{p}_A - \re \dot{\boldsymbol{\lambda}}_A \right)^T \gamma^{-\frac{1}{2}} \left[ \sum_{i = 1}^{N - n} \mathbf{v}_i \frac{1}{m_i} \mathbf{v}_i^T \right] \gamma^{-\frac{1}{2}} \left( \mathbf{p}_A - \re \dot{\boldsymbol{\lambda}}_A \right) \\
%+ \left( \mathbf{p}_A - \re \dot{\boldsymbol{\lambda}}_A \right)^T \re \dot{\boldsymbol{\lambda}}_A + \frac{1}{2} \re \dot{\boldsymbol{\lambda}}_A^T \gamma^{\frac{1}{2}} \left[ \sum_{i = 1}^{N - n} \mathbf{v}_i m_i \mathbf{v}_i^T \right] \gamma^{\frac{1}{2}} \re \dot{\boldsymbol{\lambda}}_A \\
%+ \frac{1}{2} \left( \mathbf{x}_A - \re \boldsymbol{\lambda}_A \right)^T \gamma^{\frac{1}{2}} \left[ \sum_{i = 1}^{N - n} \mathbf{v}_i \frac{\alpha_i^2}{m_i} \mathbf{v}_i^T \right] \gamma^{\frac{1}{2}} \left( \mathbf{x}_A - \re \boldsymbol{\lambda}_A \right) \\
%- \left( \mathbf{x}_A - \re \boldsymbol{\lambda}_A \right)^T \re \ddot{\boldsymbol{\lambda}}_A - \frac{1}{2} \re \dot{\boldsymbol{\lambda}}_A^T \gamma^{\frac{1}{2}} \left[ \sum_{i = 1}^{N - n} \mathbf{v}_i m_i \mathbf{v}_i^T \right] \gamma^{\frac{1}{2}} \re \dot{\boldsymbol{\lambda}}_A \\
%+ \re \dot{\boldsymbol{\lambda}}_A^T \re \dot{\boldsymbol{\lambda}}_A
%\end{multline}
%or
\begin{multline}
\hat{H}_A = \frac{1}{2} \left( \mathbf{p}_A - \re \dot{\boldsymbol{\lambda}}_A \right)^T \gamma^{-\frac{1}{2}} \left[ \sum_{i = 1}^{N - n} \mathbf{v}_i \frac{1}{m_i} \mathbf{v}_i^T \right] \gamma^{-\frac{1}{2}} \left( \mathbf{p}_A - \re \dot{\boldsymbol{\lambda}}_A \right) \\
+ \frac{1}{2} \left( \mathbf{x}_A - \re \boldsymbol{\lambda}_A \right)^T \gamma^{\frac{1}{2}} \left[ \sum_{i = 1}^{N - n} \mathbf{v}_i \frac{\alpha_i^2}{m_i} \mathbf{v}_i^T \right] \gamma^{\frac{1}{2}} \left( \mathbf{x}_A - \re \boldsymbol{\lambda}_A \right) \\
+ \mathbf{p}_A^T \re \dot{\boldsymbol{\lambda}}_A - \left( \mathbf{x}_A - \re \boldsymbol{\lambda}_A \right)^T \re \ddot{\boldsymbol{\lambda}}_A .
\label{eq:H_eff_many_coh}
\end{multline}

Let us make two special choices for the parameters $m_i$. First if we choose all $m_i$ so that
\begin{equation}
m_i = 1 ,
\end{equation}
then the effective Hamiltonian will assume the form
\begin{multline}
\hat{H}_A = \frac{1}{2} \left( \mathbf{p}_A - \re \dot{\boldsymbol{\lambda}}_A \right)^T \gamma^{-1} \left( \mathbf{p}_A - \re \dot{\boldsymbol{\lambda}}_A \right) + \mathbf{p}_A^T \re \dot{\boldsymbol{\lambda}}_A \\
+ \frac{1}{2} \left( \mathbf{x}_A - \re \boldsymbol{\lambda}_A \right)^T \left( \gamma - \beta \gamma^{-1} \beta \right) \left( \mathbf{x}_A - \re \boldsymbol{\lambda}_A \right)
- \left( \mathbf{x}_A - \re \boldsymbol{\lambda}_A \right)^T \re \ddot{\boldsymbol{\lambda}}_A .
\label{eq:H_eff_many_coh_m1}
\end{multline}

If we choose (notice that this is a time-dependent mass, and, thus, the equation \eqref{eq:H_A_Adagger_m_time} applies)
\begin{equation}
m_i = \frac{\mathbf{v}_i^T \gamma^{-\frac{1}{2}} \re \dot{\boldsymbol{\lambda}}_A}{\mathbf{v}_i^T \gamma^{\frac{1}{2}} \re \dot{\boldsymbol{\lambda}}_A} ,
\end{equation}
then the linear terms in the momentum will vanish, and the effective Hamiltonian will assume the form
\begin{multline}
\hat{H}_A = \frac{1}{2} \mathbf{p}_A^T \gamma^{-\frac{1}{2}} \left[ \sum_{i = 1}^{N - n} \mathbf{v}_i \frac{\mathbf{v}_i^T \gamma^{\frac{1}{2}} \re \dot{\boldsymbol{\lambda}}_A}{\mathbf{v}_i^T \gamma^{-\frac{1}{2}} \re \dot{\boldsymbol{\lambda}}_A} \mathbf{v}_i^T \right] \gamma^{-\frac{1}{2}} \mathbf{p}_A + \frac{1}{2} \re \dot{\boldsymbol{\lambda}}_A^T \re \dot{\boldsymbol{\lambda}}_A \\
+ \frac{1}{2} \left( \mathbf{x}_A - \re \boldsymbol{\lambda}_A \right)^T \gamma^{\frac{1}{2}} \left[ \sum_{i = 1}^{N - n} \mathbf{v}_i \frac{\alpha_i^2 \mathbf{v}_i^T \gamma^{\frac{1}{2}} \re \dot{\boldsymbol{\lambda}}_A}{\mathbf{v}_i^T \gamma^{-\frac{1}{2}} \re \dot{\boldsymbol{\lambda}}_A} \mathbf{v}_i^T \right] \gamma^{\frac{1}{2}} \left( \mathbf{x}_A - \re \boldsymbol{\lambda}_A \right) \\
- \left( \mathbf{x}_A - \re \boldsymbol{\lambda}_A \right)^T \re \ddot{\boldsymbol{\lambda}}_A .
\end{multline}
Defining $\gamma^D$ as
\begin{equation}
\gamma^D := \sum_{i = 1}^{N - n} \mathbf{v}_i \frac{\mathbf{v}_i^T \gamma^{\frac{1}{2}} \re \dot{\boldsymbol{\lambda}}_A}{\mathbf{v}_i^T \gamma^{-\frac{1}{2}} \re \dot{\boldsymbol{\lambda}}_A} \mathbf{v}_i^T ,
\end{equation}
Then, the latter assumes the form
\begin{multline}
\hat{H}_A = \frac{1}{2} \mathbf{p}_A^T \gamma^{-\frac{1}{2}} \gamma^D \gamma^{-\frac{1}{2}} \mathbf{p}_A + \frac{1}{2} \re \dot{\boldsymbol{\lambda}}_A^T \re \dot{\boldsymbol{\lambda}}_A \\
+ \frac{1}{2} \left( \mathbf{x}_A - \re \boldsymbol{\lambda}_A \right)^T \left( \gamma^{\frac{1}{2}} \gamma^D \gamma^{\frac{1}{2}} - \beta \gamma^{- \frac{1}{2}} \gamma^D \gamma^{- \frac{1}{2}} \beta \right) \left( \mathbf{x}_A - \re \boldsymbol{\lambda}_A \right) - \left( \mathbf{x}_A - \re \boldsymbol{\lambda}_A \right)^T \re \ddot{\boldsymbol{\lambda}}_A .
\end{multline}

Having traced out the subsystem $A^C$, the subsystem $A$ constitutes an open quantum system. The time evolution of such a system is non-unitary in general. Nevertheless, since \emph{all} the eigenstates of the reduced density matrix obey the equation
\begin{equation}
i \hbar \dot{\Psi}_{\left\{ n_i \right\}} = \hat{H}_A \Psi_{\left\{ n_i \right\}} ,
\end{equation}
where $\hat{H}_A$ is given by \eqref{eq:H_eff_many_coh}, equation \eqref{eq:reduced_two_eigenstates} directly implies that
\begin{equation}
i \hbar \dot{\rho}_A = \left[ \hat{H}_A , \rho_A \right] ,
\end{equation}
meaning that when the overall system lies in a coherent state, the time evolution of the reduced system \emph{is unitary} and it is driven by the Hamiltonian \eqref{eq:H_eff_many_coh}.

\subsection{The Modular Hamiltonian}

In a similar manner to the specification of the effective Hamiltonian, which evolves the reduced density matrix, we may express the modular Hamiltonian as
\begin{equation}
\hat{H}_M = - \sum_{i = 1}^{N - n} \ln \xi_i \left( \frac{1}{2 m_i \alpha_i} \left( \hat{\tilde{p}}_{Ai} - w_i \right)^2 + \frac{\alpha_i}{2 m_i} \left( \hat{\tilde{x}}_{Ai} - \tilde{x}_{0Ai} \right)^2 \right) + \frac{1}{2} \ln \xi_i - \ln \left( 1 - \xi_i \right) .
\label{eq:modular_modes}
\end{equation}
We need to revert back to the original coordinates $\mathbf{x}_{0A}$. It is a matter of tedious algebra to show that
\begin{multline}
\hat{H}_A = \frac{1}{2} \left( \mathbf{p}_A - \re \dot{\boldsymbol{\lambda}}_A \right)^T \gamma^{-\frac{1}{2}} \left[ \sum_{i = 1}^{N - n} \mathbf{v}_i \frac{\ln \xi_i}{m_i \alpha_i} \mathbf{v}_i^T \right] \gamma^{-\frac{1}{2}} \left( \mathbf{p}_A - \re \dot{\boldsymbol{\lambda}}_A \right) \\
+ \frac{1}{2} \left( \mathbf{x}_A - \re \boldsymbol{\lambda}_A \right)^T \gamma^{\frac{1}{2}} \left[ \sum_{i = 1}^{N - n} \mathbf{v}_i \frac{\ln \xi_i \alpha_i}{m_i} \mathbf{v}_i^T \right] \gamma^{\frac{1}{2}} \left( \mathbf{x}_A - \re \boldsymbol{\lambda}_A \right) \\
- \sum_{i = 1}^{N - n} \frac{1}{2} \ln \xi_i - \ln \left( 1 - \xi_i \right) .
\label{eq:H_mod_many_coh}
\end{multline}

If we choose all parameters $m_i$ to be equal to 1,
\begin{equation}
m_i = 1 ,
\end{equation}
then the modular Hamiltonian will assume the form
%\begin{multline}
%\hat{H}_A = \frac{1}{2} \left( \mathbf{p}_A - \re \dot{\boldsymbol{\lambda}}_A \right)^T \gamma^{- \frac{1}{2}} \left[ \ln \gamma^{-\frac{1}{2}} \beta \gamma^{- \frac{1}{2}} - \ln \left( I + \left( I - \gamma^{- \frac{1}{2}} \beta \gamma^{- 1} \beta \gamma^{- \frac{1}{2}} \right)^{\frac{1}{2}} \right) \left( I - \gamma^{- \frac{1}{2}} \beta \gamma^{- 1} \beta \gamma^{- \frac{1}{2}} \right)^{- \frac{1}{2}} \right] \gamma^{-\frac{1}{2}} \left( \mathbf{p}_A - \re \dot{\boldsymbol{\lambda}}_A \right) \\
%+ \frac{1}{2} \left( \mathbf{x}_A - \re \boldsymbol{\lambda}_A \right)^T \gamma^{\frac{1}{2}} \left[ \ln \gamma^{-\frac{1}{2}} \beta \gamma^{- \frac{1}{2}} - \ln \left( I + \left( I - \gamma^{- \frac{1}{2}} \beta \gamma^{- 1} \beta \gamma^{- \frac{1}{2}} \right)^{\frac{1}{2}} \right) \left( I - \gamma^{- \frac{1}{2}} \beta \gamma^{- 1} \beta \gamma^{- \frac{1}{2}} \right)^{\frac{1}{2}} \right] \gamma^{\frac{1}{2}} \left( \mathbf{x}_A - \re \boldsymbol{\lambda}_A \right) \\
%- \sum_{i = 1}^{N - n} \frac{1}{2} \ln \xi_i - \ln \left( 1 - \xi_i \right) .
%\end{multline}
\begin{multline}
\hat{H}_A = - \frac{1}{2} \left( \mathbf{p}_A - \re \dot{\boldsymbol{\lambda}}_A \right)^T K_{\mathrm{mod}} \left( \mathbf{p}_A - \re \dot{\boldsymbol{\lambda}}_A \right) \\
- \frac{1}{2} \left( \mathbf{x}_A - \re \boldsymbol{\lambda}_A \right)^T V_{\mathrm{mod}} \left( \mathbf{x}_A - \re \boldsymbol{\lambda}_A \right) \\
- \sum_{i = 1}^{N - n} \frac{1}{2} \ln \xi_i - \ln \left( 1 - \xi_i \right) ,
\label{eq:many_modular}
\end{multline}
where
\begin{align}
K_{\mathrm{mod}} &= \gamma^{- \frac{1}{2}} \left[ \ln B - \ln \left( I + \left( I - B^2 \right)^{\frac{1}{2}} \right) \right] \left( I - B^2 \right)^{- \frac{1}{2}} \gamma^{- \frac{1}{2}} , \\
V_{\mathrm{mod}} &= \gamma^{\frac{1}{2}} \left[ \ln B - \ln \left( I + \left( I - B^2 \right)^{\frac{1}{2}} \right) \right] \left( I - B^2 \right)^{\frac{1}{2}} \gamma^{\frac{1}{2}} .
\end{align}

\section{Discussion}
\label{sec:discussion}

We have managed to calculate the reduced density matrix and the modular Hamiltonian in an arbitrary harmonic system lying at an arbitrary coherent state. Furthermore, we showed that for these specific states, the time evolution of the reduced density matrix is unitary and we specified the effective Hamiltonian, which generates this unitary evolution. For a general coherent state, the reduced density matrix has non-trivial time evolution. This is natural, since the state of the composite system also evolves with time. In the special case that the composite system lies at its ground state, the state of the composite system and the reduced density matrix are time-independent. However, even in this case the system does not lie in a thermal state. Each normal mode of the reduced system actually lies in a thermal state, but the corresponding temperatures are different. These are determined by the eigenvalues $\beta^D_i$ of the matrix $B$, namely,
\begin{equation}
T_i = \frac{\alpha_i}{\ln \xi_i} ,
\end{equation}
where $\alpha_i$ and $\xi_i$ are given as functions of $\beta^D_i$ by equation \eqref{eq:a_i_xi_i}. This relation between thermodynamics and entanglement in harmonic systems at their ground state has been exploited in \cite{Eisler:2020lyn}. This temperature spectrum is not contradictory to the thermalization hypothesis \cite{Srednicki_thermalization,Deutsch}. The latter applies only to non-integrable systems.

The form of the temperatures above implies that the modular Hamiltonian, expressed in terms of the normal modes of the reduced system \eqref{eq:modular_modes} assumes the form
\begin{equation}
\hat{H}_M = - \sum_{i = 1}^{N - n} \frac{1}{T_i} \left( \frac{1}{2 m_i} \left( \hat{\tilde{p}}_{Ai} - w_i \right)^2 + \frac{\alpha_i^2}{2 m_i} \left( \hat{\tilde{x}}_{Ai} - \tilde{x}_{0Ai} \right)^2 \right) + \frac{1}{2} \ln \xi_i - \ln \left( 1 - \xi_i \right) .
\label{eq:modular_modes_T}
\end{equation}
In other words, the modular Hamiltonian contains the Hamiltonian of each normal mode of the reduced system, which appears in the effective Hamiltonian \eqref{eq:Heffective_modes}, but weighted with its inverse temperature.

The above expression of the modular Hamiltonian is naively a non-local function of the stress-energy tensor of the underlying harmonic system, since it is expressed in terms of the normal modes of the reduced system. It is well-known that in general the modular Hamiltonian is indeed a non-local function of the stress-energy tensor, except for cases with very specific characteristics. Spherical entangling surfaces in conformal field theory at the ground state is such an exceptional case \cite{Casini:2011kv}; this derivation is based on the long-known local expression of the modular Hamiltonian of the Rindler wedge \cite{Bisognano:1976za,Bisognano:1975ih}, which holds in any quantum field theory. If any of these conditions is violated, e.g. by introducing a mass term \cite{Longo:2020amm}, non-local terms emerge. The expression \eqref{eq:modular_modes_T} can be useful in the investigation of whether and under which conditions the modular Hamiltonian can be expressed as a local function of the stress-energy tensor. Such investigations have advanced more in strongly constrained systems, such as two-dimensional conformal field theory \cite{Cardy:2016fqc}.

Continuing the above discussion, the expression \eqref{eq:modular_modes_T} provides an insight of the interrelation between temperatures and the modular Hamiltonian at the level of the normal modes of the reduced system, i.e. in a non-local way. It is interesting to investigate whether this relation can be used to further probe the relation between local temperature and the modular Hamiltonian, or at least the local terms of the latter \cite{Arias:2016nip,Arias:2017dda}.

We would like to make a comment about the special case of the ground state of the composite system. In this case, the effective Hamiltonian \eqref{eq:Heffective_modes} is simply the sum of the quadratic simple harmonic oscillator Hamiltonians of each normal mode of the reduced system. The modular Hamiltonian is the same sum, but weighted with the appropriate temperatures. What the effective Hamiltonian should obey? it should provide the time evolution of the reduced density matrix, which in this case is trivial, since it is time-independent. Since the normal coordinates of the reduced system are orthogonal, any linear combination of the quadratic simple harmonic oscillator Hamiltonians of each normal mode of the reduced system commutes with the modular Hamiltonian and, thus, with the reduced density matrix. So how one chooses the effective Hamiltonian in this case? The limit of \eqref{eq:Heffective_modes} as the state of the composite system tends to the ground state is the specific selection which is continuously connected to the corresponding effective Hamiltonians that generate the dynamics of the reduced system for all classicalmost states including the ground state. So this is a natural choice for the ground state effective Hamiltonian. However, even when making this special selection, the freedom of arbitrary choice of the parameters $m_i$, results once again in this Hamiltonian being an arbitrary linear combination of the aforementioned normal modes. Notice that the freedom of these linear combination is not that innocent. The multiplication of a harmonic Hamiltonian with a constant greatly alters the dynamics it describes as it alters the eigenfrequency of the corresponding oscillator.

There is also another property of the spectrum of the reduced density matrix, which complicates the calculation of the modular Hamiltonian, via the formula \eqref{eq:modular_modes}. In the case that the subsystem $A^C$ has fewer degrees of freedom than subsystem $A$, the spectrum of the matrix $B$ contains at least as many vanishing eigenvalues as the difference of the numbers of the degrees of freedom of the two subsystems. It is simple to prove this statement. Performing a Schmidt decomposition of the state of the composite system state yields
\begin{equation}
\left| \Psi \right> = \sum_{i = 1}^m \sum_{j = 1}^m c_i \left| \psi_{Ai} \right> \otimes \left| \psi_{A^C i} \right> ,
\end{equation}
where the states $\left| \psi_{Ai} \right>$ and $\left| \psi_{A^C i} \right>$ are sets of orthonormal states of the subsystems $A$ and $A^C$, which contain equal number of states at most equal to the dimensionality of the smaller subsystem, i.e.
\begin{equation}
m \leq \min \left( \dim A , \dim A^C \right) .
\end{equation}
This directly implies that
\begin{equation}
\rho_A = \sum_{i = 1}^m c_i c^*_i \left| \psi_{Ai} \right> \left< \psi_{Ai} \right| , \quad \rho_{A^C} = \sum_{i = 1}^m c_i c^*_i \left| \psi_{A^C i} \right> \left< \psi_{A^C i} \right| .
\end{equation}
It directly follows that the two reduced density matrices have exactly the same spectrum up to vanishing eigenvalues. These vanishing eigenvalues correspond to the vectors of the Hilbert space of the larger subsystem, which are orthogonal to the vectors $\left| \psi_{Ai} \right>$ or $\left| \psi_{A^C i} \right>$, depending on which is the larger subsystem.

This means that the larger subsystem necessarily contains modes which are disentangled to the smaller subsystem, and, thus, remain frozen upon tracing out the smaller subsystem. Equation \eqref{eq:modular_modes_T} directly implies that the modular Hamiltonian of the larger subsystem diverges. This apparent problem can be resolved via the introduction of an IR cutoff such as temperature. If the composite system lies at a thermal state instead of its ground state, these vanishing temperatures will tend to become equal to the temperature of the composite system yielding the modular Hamiltonian finite.

As a final comment, in the special case that all temperatures $T_i$ become equal to each other, the quadratic parts of the effective Hamiltonian and the modular Hamiltonian become proportional to each other. In the special case of the ground state there are no linear terms and the two Hamiltonians become literally proportional to each other.

These results have a profound application in the spirit of \cite{srednicki}. In the seminal work of Srednicki, free scalar field theory is treated as a harmonic system upon discretization of the degrees of freedom. The determination of the corresponding modular Hamiltonian in this case may provide more hints for the similarity between entanglement and gravity beyond the similarity of entanglement and black hole entropies. More specifically, we propose that the gravitational dynamics may be encoded in the structure of the modular Hamiltonian. The modular Hamiltonian obeys the first law of entanglement thermodynamics with respect to entanglement entropy. The latter is dominated by an area law as shown in the same work \cite{srednicki}. As a direct consequence of \cite{Jacobson_old} the modular Hamiltonian should somehow embody Einstein gravity dynamics. Having written down the modular Hamiltonian in the form \eqref{eq:many_modular} allows to check whether the above argument indeed implies that the modular Hamiltonian resembles the discretized Hamiltonian for scalar degrees of freedom propagating on a curved background, greatly enhancing the understanding of the relation between gravity and entanglement.

\section*{Acknowledgements}

D. K. was supported by FAPESP Grant No. 2021/01819-0. The authors would like to thank K. Papadodimas for useful discussions.

\appendix
\renewcommand{\thesection}{\Alph{section}}
\numberwithin{equation}{section}
\section{Coherent and Squeezed States of the Simple Harmonic Oscillator}
\label{sec:squeezed}

A simple application of the formulation of the generalized, time-dependent oscillator that we developed in Section \ref{sec:Osc} is the construction of the Gaussian states of the simple harmonic oscillator, namely the coherent and squeezed states \cite{squeezed_original}. Although this material is well-known and easily accessible in the literature, we present it for completeness and in order to demonstrate how easily it emerges in this formalism.

In Section \ref{sec:Osc} we showed that the Gaussian state \eqref{eq:Psi_0} solves the time-dependent Schrödinger equation with Hamiltonian given by \eqref{eq:H_x_p}. This Hamiltonian coincides with the Hamiltonian of the simple harmonic oscillator with eigenfrequency equal to $\omega$, provided that
\begin{align}
\omega^I &= \frac{\dot{\omega}^R}{2 \omega^R} , \label{eq:ar_dif_eq0}\\
\omega^2 &= \frac{d}{dt} \left( \frac{\dot{\omega}^{R}}{2 \omega^R} \right) - \left( \frac{\dot{\omega}^{R}}{2 \omega^R} \right)^2 + \left( \omega^R \right)^2 , \label{eq:ar_dif_eq} \\
p_0 &= m \dot{x}_0 , \quad \ddot{x}_0 + \omega^2 x_0 = 0 . \label{eq:ar_dif_eq2}
\end{align}
The equations \eqref{eq:ar_dif_eq2} are solved by
\begin{equation}
x_0 = \alpha \cos \left( \omega t + \phi_0 \right) , \qquad p_0 = - m \omega \alpha \sin \left( \omega t + \phi_0 \right) .
\end{equation}

In order to solve equation \eqref{eq:ar_dif_eq}, we set
\begin{equation}
\omega^R = \frac{\omega}{f^2} .
\end{equation}
Then, $f$ obeys the equation
\begin{equation}
\frac{1}{\omega^2} \ddot{f} + f - \frac{1}{f^3} = 0 .
\end{equation}
We can integrate this equation once to obtain
\begin{equation}
\frac{1}{4 \omega^2}\left[ \frac{\partial}{\partial t} \left( f^2 \right) \right]^2 = - f^4 + 2 \cosh \gamma f^2 - 1 ,
\end{equation}
where we defined the integration constant as $2 \cosh \gamma$. The integration constant has been selected so that the right-hand-side of the above equation can be positive, allowing the existence of a real solution. For this purpose, the integration constant has to be larger than 2, and thus we defined it as $2 \cosh \gamma$. Solving this equation yields
\begin{equation}
f^2 = \cosh \gamma + \sin \left( 2 \omega \left( t - t_0 \right) \right) \sinh \gamma ,
\end{equation}
implying that $\omega^R$ is given by
\begin{equation}
\omega^R = \frac{\omega}{\cosh \gamma + \sin \left( 2 \omega \left( t -t_0 \right) \right) \sinh \gamma} .
\end{equation}
Equation \eqref{eq:ar_dif_eq0} directly implies that
\begin{equation}
\omega^I = - \frac{\omega \cos \left( 2 \omega \left( t - t_0 \right) \right) \tanh \gamma}{1 + \sin \left( 2 \omega \left( t - t_0 \right) \right) \tanh \gamma} .
\end{equation}
Finally, the phase $\phi$, which appears in the wavefunctions \eqref{eq:wavefunctions_general}, is simply the time integral of $\omega^R$, as given by \eqref{eq:time_phase}. It turns out that it is equal to
\begin{equation}
\phi = \arctan \left[ \frac{\tan \left( \omega \left( t - t_0 \right) \right) + \tanh \frac{\gamma}{2}}{1 + \tan \left( \omega \left( t - t_0 \right) \right) \tanh \frac{\gamma}{2}} \right] .
\end{equation}
The potential of the generalized harmonic oscillator is not exactly the potential of the simple harmonic oscillator; it contains a c-number term too. This term is eliminated by
\begin{equation}
\Psi_n \left( x , t \right) \rightarrow e^{- i \frac{m \omega a^2}{4 h} \sin \left( 2 \omega t + \phi \right)} \Psi_n \left( x , t \right) .
\end{equation}
Thus, the time-dependent Schrödinger equation
\begin{equation}
i \hbar \frac{\partial}{\partial t} \Psi = - \frac{\hbar^2}{2 m} \frac{\partial^2}{{\partial x}^2} \Psi + \frac{1}{2} m \omega^2 x^2 \Psi ,
\end{equation}
has the following family of solutions
\begin{multline}
\Psi_n = \frac{1}{\sqrt{2^n n!}} \left(\frac{m \omega^R}{\pi \hbar} \right)^{\frac{1}{4}}H_n \left( \sqrt{\frac{m \omega^R}{\hbar}} \left( x - x_0 \right) \right) \\
\times \exp \left[ - \frac{m \left( \omega^R + i \omega^I \right)}{2 \hbar} \left( x - x_0 \right)^2 + i \frac{p_0}{\hbar} \left( x - x_0 \right) - i \phi_n \right] ,
\end{multline}
where
\begin{equation}
\phi_n = \left( n + \frac{1}{2} \right) \arctan \left[ \frac{\tan \left( \omega \left( t - t_0 \right) \right) + \tanh \frac{\gamma}{2}}{1 + \tan \left( \omega \left( t - t_0 \right) \right) \tanh \frac{\gamma}{2}} \right] + \frac{m \omega}{4 h} \alpha^2 \sin \left( 2 \left( \omega t + \phi \right) \right) .
\end{equation}
Notice, that the $n = 0$ wavefunction is Gaussian and actually it describes a squeezed state. In the special case $\gamma = 0$, this reduces to a coherent state. 

Using the usual creation and annihilation operator techniques, it is straightforward to calculate the following expectation values
\begin{align}
\langle x \rangle &= x_0 , \\
\langle x^2 \rangle &= \left( n + \frac{1}{2} \right) \frac{\hbar}{m \omega^R} + x_0^2 , \\
\langle p \rangle &= p_0 \\
\langle p^2 \rangle &= \left( n + \frac{1}{2} \right) \hbar m \omega^R \left( 1 + \left( \frac{\omega^I}{\omega^R}\right)^2 \right) + p_0^2 ,
\end{align}
when the system is described by the wave-functions \eqref{eq:wavefunctions_general}. Thus, we obtain
\begin{equation}
\Delta x \Delta p = \hbar \left( n + \frac{1}{2} \right) \sqrt{1 + \left( \frac{\omega^I}{\omega^R} \right)^2} .
\end{equation}
In the case of squeezed states ($n = 0$), it follows that
\begin{equation}
\Delta x \Delta p = \frac{\hbar}{2} \sqrt{1 + \cos^2 \left( 2 \omega \left( t - t_0 \right) \right) \sinh^2 \gamma} ,
\end{equation}
which is the well-known uncertainty relation of the squeezed states.

Summing up, the algebraic construction of the generalized time-dependent oscillator, which is presented in Section \ref{sec:Osc}, trivializes the construction of the Gaussian states of the simple harmonic oscillator.

\end{document}